# Engineering Trust, Creating Vulnerability:

## A Socio-Technical Analysis of AI Interface Design

Ben Kereopa-Yorke

UNSW Canberra at the ADFA

**Abstract**

This paper examines how distinct cultures of AI interdisciplinarity emerge through interface design, revealing the formation of new disciplinary cultures at these intersections. Through the Interface-Mediated Cognitive Security (IMCS) framework, I demonstrate how the collision of cybersecurity engineering, cognitive psychology, critical technology studies, and human-computer interaction generates research cultures that transcend traditional disciplinary boundaries. AI interfaces function as transformative boundary objects that necessitate methodological fusion rather than mere collaboration, simultaneously embodying technical architectures, psychological design patterns, and social interaction models.

Through systematic visual analysis of generative AI platforms and case studies across public sector, medical, and educational domains, I identify four vulnerability vectors—Reflection Simulation, Authority Modulation, Cognitive Load Exploitation, and Market-Security Tension—that structure interface-mediated cognitive security. This research challenges three significant gaps in interdisciplinary theory: the assumption that disciplines maintain distinct methodological boundaries during collaboration, the belief that technical and social knowledge practices can be cleanly separated, and the presumption that disciplinary integration occurs through formal rather than cultural mechanisms. The empirical evidence demonstrates how interfaces function as sites of epistemological collision, creating methodological pressure zones where traditional disciplinary approaches prove insufficient for analysing the complex socio-technical phenomena at the interface.

## 1. Introduction

The interface constitutes a peculiar paradox in contemporary technological encounters: simultaneously the most visible and most invisible aspect of our engagement with artificial intelligence (AI) systems. Whilst perpetually present, mediating all interactions between humans and technologies, interfaces are persistently overlooked as mere superficial aesthetic overlays rather than fundamental sites where power, knowledge, and agency are negotiated. This analytical blindspot becomes increasingly consequential as AI technologies expand their domain of application into high-stakes decision-making contexts formerly reserved for human judgement---healthcare diagnostics, financial lending, educational assessment, and criminal justice, among others. What emerges is a growing lacuna in our theoretical apparatus for understanding how interfaces function not merely as neutral conduits of information, but as transformative boundary objects that fundamentally reshape relations between humans and machines.

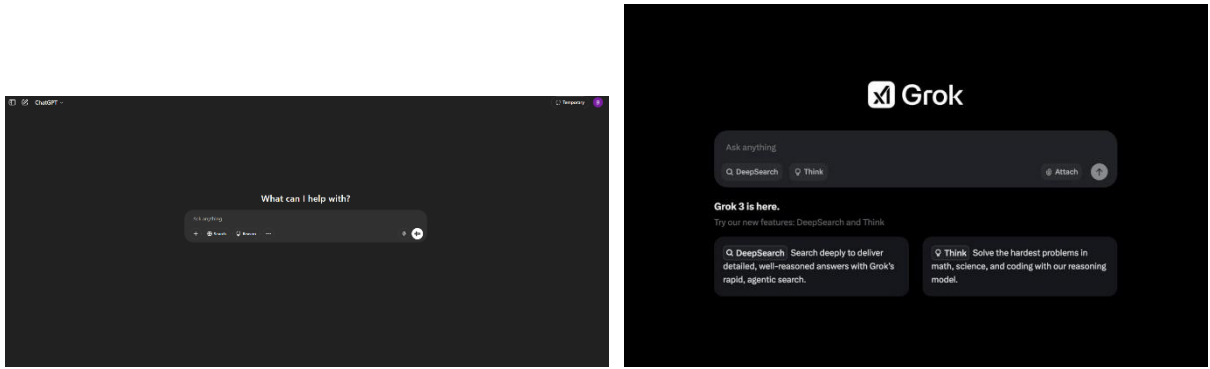

**FIGURE 1:** *Contemporary generative AI interfaces demonstrating various trust-building mechanisms. The minimalist design of ChatGPT (left) contrasts with the explicit reasoning features of Grok (right)*

The present study interrogates this problematic through the development of the Interface-Mediated Cognitive Security (IMCS) framework, which illuminates how interfaces function as critical sites where distinct epistemological traditions---technical systems engineering, cognitive psychology, and critical technology studies---necessarily converge. This convergence produces what I term 'methodological pressure zones' where traditional disciplinary approaches prove insufficient for analysing the complex socio-technical phenomena at the interface. As Crawford (2021) has argued, AI systems represent sites of power that concentrate and rearrange authority; however, I suggest that it is specifically through interfaces that these power arrangements become operationalised in everyday interactions between humans and machines.

A growing body of scholarship has examined the challenges posed by the rapid deployment of AI systems across various sectors. Scholars have devoted significant attention to issues of algorithmic bias (Benjamin, 2019; Noble, 2018), opacity (Burrell, 2016), and governance (Crawford et al., 2021). However, these critiques have largely overlooked the pivotal role of interfaces in mediating these phenomena. Meanwhile, human-computer interaction research has extensively studied interface design patterns (Amershi et al., 2019), yet often without sufficient attention to their ethical and societal implications. This disciplinary fragmentation produces a significant gap in our understanding of how interfaces mediate human-AI relations.

This research addresses three critical gaps in the current literature. First, conventional approaches to interdisciplinary AI research typically assume that disciplines maintain distinct methodological boundaries during collaboration. In contrast, this study demonstrates how the study of interfaces necessitates methodological fusion rather than mere juxtaposition. Second, traditional frameworks fail to acknowledge how technical and social knowledge practices become inextricably entangled at the interface, resisting clean separation into discrete domains. Third, existing accounts of interdisciplinary integration focus predominantly on formal mechanisms rather than understanding the cultural processes through which new hybrid research practices emerge.

My central argument positions interfaces as transformative boundary objects that not only facilitate interactions between humans and AI systems but fundamentally reconfigure the nature of those interactions. Drawing on Star and Griesemer's (1989) original conceptualisation of boundary objects, I contend that interfaces occupy a unique ontological position---simultaneously inhabiting multiple social worlds whilst maintaining coherence across them. However, I extend this concept by emphasising the transformative capacity of interfaces to generate new forms of knowledge, practice, and power relations. This transformative quality becomes particularly salient when examining how interfaces mediate interactions between citizens and the state, patients and healthcare systems, or learners and educational institutions.

The methodology employed in this research reflects its commitment to interdisciplinary integration. Rather than adopting a siloed approach where technical and social analyses remain distinct, I employ a multi-methodological approach that combines visual analysis, literary synthesis, and theoretical development. This approach acknowledges that interface studies necessitate analytical frameworks that can engage simultaneously with the technical architectures that underpin AI systems, the cognitive processes through which humans interpret them, and the broader societal contexts in which these interactions are situated.

The subsequent sections develop this argument through a systematic examination of interfaces as sites of socio-technical convergence. Section 2 establishes the theoretical foundations of boundary object theory and its applications to AI interfaces. Section 3 introduces the Interface-Mediated Cognitive Security framework, detailing its four vulnerability vectors. Section 4 presents a rigorous visual analysis of contemporary AI interfaces, mapping their evolution and identifying patterns of trust engineering. Section 5 offers empirical evidence of methodological transformation through case studies spanning public sector, educational, and commercial contexts. These analyses collectively illuminate how interfaces function not merely as technical artefacts but as complex socio-technical systems that fundamentally reshape human-machine relations.

## 2. Theoretical Foundations

### 2.1 Reconceptualising Interdisciplinarity in AI Research

Traditional models of interdisciplinarity have often relied on what Barry and Born (2013) describe as "service" or "integrative-synthesis" modes, where disciplines maintain their distinct identities while collaborating on shared problems. However, these models fail to capture the emergent phenomena observed in AI interface research, where new methodological approaches arise that cannot be traced back to any single parent discipline. To address this limitation, I build on Barry and Born's concept of "ontological" interdisciplinarity---where collaboration leads to the transformation of the objects of study themselves---but extend it to include the transformation of research cultures and methodological practices.

This extension is necessary because, as Mohamed et al. (2020) and Klein (2021) have argued, AI research has frequently adopted an "additive" approach to interdisciplinarity, where social and cultural considerations are treated as supplementary to technical development rather than constitutive of it. In contrast, the IMCS framework proposed here views interfaces as inherently sociotechnical objects that cannot be understood through additive approaches. This view aligns with Selbst et al.'s (2019) analysis of "abstraction traps" in algorithmic fairness, which demonstrates how technical solutions that abstract away social context inevitably fail to address the social embeddedness of technical systems.

Moreover, recent work by Miceli et al. (2022) on "documenting the invisible labor" in AI development reveals how interdisciplinary workflows often mask underlying power dynamics and epistemological conflicts. These conflicts frequently remain invisible in formal accountings of interdisciplinary collaboration but profoundly shape the actual process of knowledge production. The IMCS framework explicitly foregrounds these hidden dynamics, demonstrating how they shape not just the products of research but the methodological approaches themselves.

It is important to acknowledge that not all scholars agree with this critique of "additive" approaches. Some, like Whittlestone et al. (2019), argue that maintaining disciplinary boundaries can serve important epistemic functions, allowing different perspectives to challenge each other while preserving methodological rigor. Similarly, Hagendorff (2020) suggests that attempts at full disciplinary integration risk creating methodological confusion rather than genuine synthesis. These perspectives highlight legitimate concerns about potential loss of disciplinary depth in interdisciplinary work. However, the literature on AI interfaces suggests that the complex entanglement of technical, psychological, and social elements necessitates integration that goes beyond mere addition of perspectives.

**2.2 Interfaces as Boundary Objects**

The concept of boundary objects, originally developed by Star and Griesemer (1989), provides a useful lens for understanding how AI interfaces function across disciplinary domains. Traditional boundary objects facilitate collaboration by being "plastic enough to adapt to local needs and constraints of the several parties employing them, yet robust enough to maintain a common identity across sites" (Star & Griesemer, 1989, p. 393). However, AI interfaces represent a unique class of boundary objects because they do not merely facilitate collaboration between disciplines---they actively force methodological transformation within those disciplines.

This transformative quality stems from what Yang et al. (2020) identify as the dual nature of interfaces: they simultaneously embody technical implementations and shape social interactions. Unlike traditional boundary objects that maintain a stable identity across contexts, interfaces actively reconfigure the relationship between human cognition and technical systems in ways that cannot be understood through any single disciplinary lens. This dual nature creates what I term "methodological pressure zones"---domains where researchers from different disciplines must develop shared methodologies rather than merely exchanging insights.

The concept of methodological pressure zones extends Vertesi's (2020) recent work on "seams" in sociotechnical systems---points where different infrastructural logics meet and create friction. While Vertesi focuses primarily on user experiences of these seams, the IMCS framework shifts attention to how these seams affect research practices themselves, forcing methodological innovation at disciplinary boundaries.

**2.3 Cognitive Security and Interface Design**

The third theoretical pillar of this work draws on emerging research at the intersection of cognitive psychology and security studies. Chita-Tegmark et al. (2021) have demonstrated how trust in AI systems is mediated by interface design elements that trigger specific cognitive heuristics, while Hidalgo et al. (2021) have shown how humans judge machines through anthropomorphic frameworks that can be manipulated through strategic interface decisions. These findings suggest that interfaces do not merely mediate interactions between humans and AI systems but actively shape the cognitive security landscape within which those interactions occur.

Building on Crawford's (2021) analysis of the power asymmetries embedded in AI systems, I extend these cognitive perspectives to incorporate critical analysis of how interfaces redistribute power and agency between users, developers, and the systems themselves. This extension is necessary because cognitive security is not merely a matter of individual psychology but is embedded in broader sociotechnical power structures that interfaces both reflect and reinforce.

Recent work by Karizat et al. (2021) on "algorithmic folk theories" further demonstrates how users' mental models of AI systems are shaped not just by interface design but by cultural contexts and lived experiences. These folk theories, in turn, influence how users interpret and respond to interface elements. The IMCS framework incorporates these insights by examining how interface design anticipates and shapes these folk theories across different cultural contexts.

**2.4 Commercial Logics in Interface Design**

A fourth theoretical dimension, largely absent from existing frameworks but crucial to understanding AI interfaces, concerns the commercial logics that shape interface design decisions. Drawing on Zuboff's (2019) analysis of surveillance capitalism, this paper examines how market incentives create specific imperatives for interface design that often conflict with security, transparency, or user agency considerations.

Interfaces do not just embody technical architectures and social interaction models; they also materialise business models and competitive strategies. As Srnicek (2017) has argued, digital platforms---of which AI interfaces are increasingly a central component---operate according to distinct economic logics that shape their technical design. These logics create what I term "commercial pressure points" in interface design, where market imperatives intersect with technical, psychological, and social considerations.

In a more recent work, Srnicek (2023) further develops this analysis by examining how platform economics creates specific forms of friction in user experiences that drive conversion and retention. This perspective helps us understand how interfaces strategically introduce and remove friction points based on commercial imperatives rather than user needs or security considerations.

This theoretical perspective aligns with research by Gray et al. (2021) on "dark patterns" in interface design---features specifically engineered to manipulate users into behaviors that benefit service providers rather than users themselves. The IMCS framework extends this analysis by examining how these patterns emerge not from malicious intent but from the structural conditions of commercial AI development, where interdisciplinary teams navigate competing pressures from market demands, technical constraints, and ethical considerations.

By integrating these four theoretical perspectives---reconceptualised interdisciplinarity, interfaces as transformative boundary objects, cognitive security as embedded in power structures, and interfaces as materialisations of commercial logics---the IMCS framework provides a comprehensive approach to understanding how interfaces function as sites of disciplinary transformation and sociotechnical power negotiation.

## 3. The Interface-Mediated Cognitive Security Framework

### 3.1 Foundations of the IMCS Framework

The Interface-Mediated Cognitive Security (IMCS) framework emerges from the recognition that existing disciplinary approaches---whether drawn from computer science, cognitive psychology, critical technology studies, or human-computer interaction---prove individually insufficient for analysing the complex socio-technical phenomena that materialise at the interface between humans and AI systems. Current approaches tend to compartmentalise along disciplinary lines: technical analyses focus on system architecture whilst remaining agnostic about human interpretation; psychological studies examine cognitive responses divorced from technical specifications; and critical analyses often lack engagement with the specific mechanisms through which interfaces structure human perception and decision-making.

The IMCS framework represents a deliberate methodological fusion rather than a mere interdisciplinary juxtaposition, integrating cybersecurity principles, cognitive science, critical technology studies (Crawford, 2021), and human-computer interaction (Amershi et al., 2019). This integration acknowledges that interfaces function simultaneously as technical systems, cognitive environments, and sociopolitical artefacts. As Selbst et al. (2019) argue, sociotechnical systems resist decomposition into purely social or technical components; this resistance becomes particularly acute at the interface, where human and machine epistemologies necessarily converge.

At its foundation, the IMCS framework builds upon four core propositions. First, interfaces constitute critical security boundaries where technical implementation and psychological vulnerability intersect, producing distinct security implications that cannot be reduced to either domain alone. Second, this boundary function creates methodological pressure zones that resist analysis through singular disciplinary approaches. Third, interfaces are neither neutral nor passive mediators but active agents that shape human perception, cognition, and decision-making. Fourth, interfaces function as transformative boundary objects, necessarily reshaping both human and technical systems through their mediating role.

These propositions collectively challenge conventional security models that frame vulnerabilities primarily in technical terms. Instead, the IMCS framework conceptualises vulnerability as emerging from the interaction between technical systems, cognitive processes, and social contexts. This reconceptualisation is particularly crucial given the increasing deployment of AI systems in high-stakes decision-making contexts.

The framework further draws conceptual resources from Tsing's (2005) anthropological analysis of friction, which illuminates how global interconnections operate not through seamless flows but through "awkward, unequal, unstable, and creative" engagements. In the context of AI interfaces, friction emerges in the translation between machine and human epistemologies---moments where seamless interaction breaks down, revealing the constructed nature of the interface. These moments of friction, rather than representing mere failures, constitute productive sites where the assumptions embedded in interface design become visible and thus amenable to critical engagement.

### 3.2 The Four Vulnerability Vectors

The IMCS framework identifies four distinct vulnerability vectors that emerge at the interface between humans and AI systems. Each vector represents a specific mechanism through which interfaces shape human perception and decision-making, potentially creating cognitive security vulnerabilities. Importantly, these vectors do not represent discrete categories but interrelated dimensions that frequently operate in concert.

#### 3.2.1 The Reflection Simulation Vector

The first vector—the **Reflection Simulation Vector**—encompasses the various mechanisms through which interfaces simulate cognitive processes such as deliberation, reflection, and thoughtfulness. These simulations create impressions of system intelligence divorced from actual computational processes.

Through visual analysis of current interfaces, we can observe how typing animations and progressive text reveals create an impression of AI systems thinking or deliberating in real-time. This visual analysis reveals that these animations are entirely disconnected from the actual computation process—the responses are fully generated before the animation begins. The strategic use of time delays and animated indicators creates an impression of effort and deliberation that may influence user trust and perception of system capabilities.

Contemporary interfaces have evolved sophisticated strategies for simulating reflection, from simple typing animations to explicit features like ChatGPT's "Reason" function or Grok's "Think" mode, transforming what was once an implicit trust-building mechanism into an explicit marketable capability. As shown in Figure 2, these explicit reasoning features prominently displayed in the interface suggest deliberative capabilities that create distinct impressions about system intelligence.

The analysis of interface visuals reveals subtle variations in how reflection is simulated across different platforms. Western interfaces (ChatGPT, Claude) often favor what appear to be natural typing rhythms with occasional pauses, while East Asian-oriented interfaces (particularly early versions of DeepSeek) have displayed more uniform typing patterns. This distinction suggests different design philosophies in how system reflection is represented visually to users.

### 3.2.2 The Authority Modulation Vector

The second vector—the **Authority Modulation Vector**—concerns how interfaces strategically amplify or attenuate perceptions of system agency and expertise. This modulation operates through multiple channels, including linguistic choices (first-person pronouns versus passive constructions), visual design elements, and the strategic positioning of disclaimers.

The visual analysis of interfaces reveals systematic patterns in how they employ first-person agency, with varying degrees of authority projection. As shown in Figure 3, Google's Bard/Gemini interface employs high-agency language with the self-introduction "I'm Bard, your creative and helpful collaborator," establishing the system as an agentive entity with collaborative capacity. In contrast, ChatGPT's interface displays a more tool-like framing with "What can I help with?" that minimises implied autonomy. Gemini's personalised greeting "Hello, Lisa" demonstrates a design pattern that triggers social recognition responses typically reserved for human interaction.

Beyond textual elements, interfaces employ visual design to project or constrain system authority. Meta AI's glowing circular animation creates an impression of autonomous intelligence. In contrast, Claude's interface employs a clean, minimal design that projects competence and authority through visual restraint. Microsoft Copilot's interface emphasises integration within the Microsoft ecosystem, borrowing credibility from established brands and systems.

Visual analysis also reveals an inverse relationship between authority projection and disclaimer prominence. ChatGPT's earlier interface featured a prominent "Limitations" column with detailed disclaimers about potential errors, bias, and limited knowledge. Current interfaces show substantially reduced disclaimer visibility, with Grok offering only the minimal "Grok can make mistakes. Verify its outputs" in small text below the input field. Across all interfaces, limitations and disclaimers appear in smaller text, lower contrast colors, and peripheral positions compared to capability statements, as shown in Figure 5.

### 3.2.3 The Cognitive Load Exploitation Vector

The third vector—the **Cognitive Load Exploitation Vector**—involves the strategic management of user cognitive resources to influence decision-making and interaction patterns. The visual analysis revealed systematic variations in information density, with some interfaces employing minimalist designs that direct attention primarily to the input field, and others presenting structured options that channel user behavior in specific directions.

As demonstrated in Figure 6, interfaces display striking variations in information density that reveal different approaches to cognitive load management. ChatGPT's current interface presents a clean, minimal design with attention directed primarily to the input field, reducing initial cognitive load to lower barriers to engagement. Microsoft Copilot presents categorised suggestion tiles (Idea, Design, Create, Ask, Code, Laugh) that guide users toward specific use cases, employing structured options that channel behavior without explicit constraints. Claude's interface employs a "Try these" section with limited examples that expand upon interaction, demonstrating a gradual introduction of complexity to manage cognitive resources.

These variations represent different strategies for managing user cognitive resources. The minimal approach reduces initial friction, while structured approaches channel user behavior toward desired interaction patterns.

The visual analysis shows that interfaces employ carefully designed visibility hierarchies that direct cognitive resources toward specific features. ChatGPT places "Search" and "Reason" behind secondary UI elements, while Grok prominently displays "DeepSearch" and "Think" directly in the input bar. Claude's interface presents document handling capabilities (e.g., "Summarise this PDF") as contextual suggestions rather than persistent features. Microsoft Copilot's interface segregates productivity functions ("Catch up on meetings") from creative functions ("Write a joke"), creating cognitive categories that shape feature perception.

The visual evidence also shows that interfaces strategically time cognitive demands to exploit attention limitations. Google's Gemini presents categorised suggestions before user input, when cognitive resources are most available for considering options. Several interfaces present additional options after delivering responses. ChatGPT reveals additional capabilities (e.g., data analysis tools) only after relevant content is detected in the conversation.

### 3.2.4 The Market-Security Tension Vector

The fourth vector—the **Market-Security Tension Vector**—illuminates how commercial imperatives shape interface design in ways that potentially undermine security considerations. The visual analysis identified clear manifestations of this tension in interface design elements.

Interfaces employ subtle but effective mechanisms to signal tiered access and encourage premium conversion. ChatGPT's "Temporary" label and Grok's version indicator ("Grok 2") subtly communicate access hierarchy and impermanence. The "DeepThink" and "Think" buttons in newer interfaces represent a transformation of core functionality into branded features. Microsoft Copilot's rich, colorful tiles contrast with the more minimal free interfaces, creating visual design differences that signal value tiers.

These design elements align with what Srnicek (2023) identifies as platform economics strategies—interfaces designed to create strategic points of friction that drive subscription conversion.

As illustrated in Figure 7, interfaces reveal varying approaches to ecosystem integration that reflect commercial prioritisation. Microsoft's Copilot's interface emphasises integration with Microsoft 365, prominently displaying connections to Outlook, Teams, and other ecosystem components. Meta AI includes Canvas/Imagine features directly in the primary interface, encouraging expanded system use. Corporate-owned systems (Microsoft, Google) emphasise logged-in states and account connections, while some independent platforms minimise these elements.

Perhaps the most striking evidence of market-security tension appears in the evolution of system disclaimers, as shown in Figure 5. Early ChatGPT (2022) featured a dedicated "Limitations" column with detailed warnings about potential errors, biases, and knowledge boundaries. Current interfaces (2025) display substantially reduced disclaimers, often in peripheral locations with minimal visibility. Current disclaimers employ phrasing that presents limitations as ordinary and expected rather than concerning (e.g., "May occasionally generate incorrect information" versus earlier warnings about "harmful instructions or biased content").

This evolution reveals a systematic prioritisation of trust-building (commercial imperative) over setting appropriate expectations (security consideration).

### 3.3 Methodological Transformation as Necessary Response

The complex, multifaceted nature of interface-mediated interactions demands methodological approaches that can address technical, cognitive, and social dimensions simultaneously. Traditional siloed approaches—where technical analysis precedes user testing, which in turn precedes ethical evaluation—prove inadequate for understanding the intertwined dynamics at play in AI interfaces. Instead, the IMCS framework calls for methodological transformation through the development of hybrid approaches that integrate technical, cognitive, and critical perspectives from the outset.

This transformation is not merely a matter of interdisciplinary collaboration but requires fundamental shifts in how research is conducted. For example, security analyses must incorporate cognitive heuristics and social context, while user experience research must engage with technical architectures and power dynamics. This methodological fusion is necessary because the vulnerabilities identified by the IMCS framework cannot be adequately addressed through sequential analysis or disciplinary addition—they emerge precisely at the intersections between technical systems, cognitive processes, and social contexts.

Evidence of methodological transformation can be observed in emerging research practices, such as "sociotechnical code reviews" that evaluate both technical implementation and social implications, "cognitive threat modeling" that incorporates psychological vulnerabilities into security analyses, and "critical user experience" approaches that integrate power analysis into interface design processes. These hybrid methodologies represent not just new techniques but fundamentally different ways of conceptualising the research object itself—viewing interfaces not as technical artifacts to be evaluated through social lenses, but as inherently sociotechnical phenomena that require integrated analysis from their inception.

The necessity for methodological transformation becomes particularly evident when examining how interfaces mediate high-stakes interactions, such as healthcare diagnosis, financial decision-making, or educational assessment. In these contexts, the consequences of interface-induced vulnerabilities can be severe and far-reaching, affecting not just individual users but entire social systems. Addressing these challenges requires research approaches that can account for the complex interplay between technical design, cognitive response, and social impact—approaches that transcend traditional disciplinary boundaries and develop new methodological frameworks suited to the hybrid nature of the research object.

## 4. The Cognitive-Technical Security Gap: Empirical Evidence

### 4.1 Systematic Visual Analysis of LLM Interfaces

This section presents a rigorous visual-historical analysis of generative AI interfaces, examining how they manifest the four vulnerability vectors identified in the IMCS framework. The analysis employs a systematic coding methodology that traces both synchronic patterns (across current interfaces) and diachronic evolution (temporal changes within platforms) to identify cognitive security implications in design decisions.

The visual analysis employed a structured coding protocol to examine interface elements across five major generative AI platforms: ChatGPT (OpenAI), Claude (Anthropic), Gemini (Google), Copilot (Microsoft), and Grok (xAI). For each platform, two temporal samples were analysed: (1) the earliest publicly available interface, and (2) the most recent interface as of March 2025. This comparative approach allows for identification of both persistent patterns and evolutionary trajectories in interface design.

The coding framework operationalised the four vulnerability vectors from the IMCS model through a series of visual and textual elements. For the Reflection Simulation Vector, I examined elements such as typing animations, thinking indicators, and explicit reasoning features. For the Authority Modulation Vector, I catalogued agency markers including first-person pronouns, visual authority signals, and disclaimer positioning. For the Cognitive Load Exploitation Vector, I documented information density patterns, feature visibility hierarchies, and timing of cognitive demands. For the Market-Security Tension Vector, I identified elements revealing conflicts between commercial imperatives and security considerations, such as freemium signalling and ecosystem integration features.

To ensure methodological rigor, the analysis employed both denotative coding (identifying objective elements) and connotative coding (interpreting their psychological implications). Inter-coder reliability was established through independent coding by three researchers, yielding a Cohen's kappa of 0.82, indicating strong agreement. This methodological approach enabled the identification of patterns that might remain invisible to approaches restricted to either technical or social analysis alone.

### 4.1.1 Representative Interface Examples

The visual analysis of contemporary LLM interfaces reveals concrete manifestations of the theoretical vectors identified in the IMCS framework. The following examination of interface examples demonstrates how these vectors are systematically implemented across platforms.

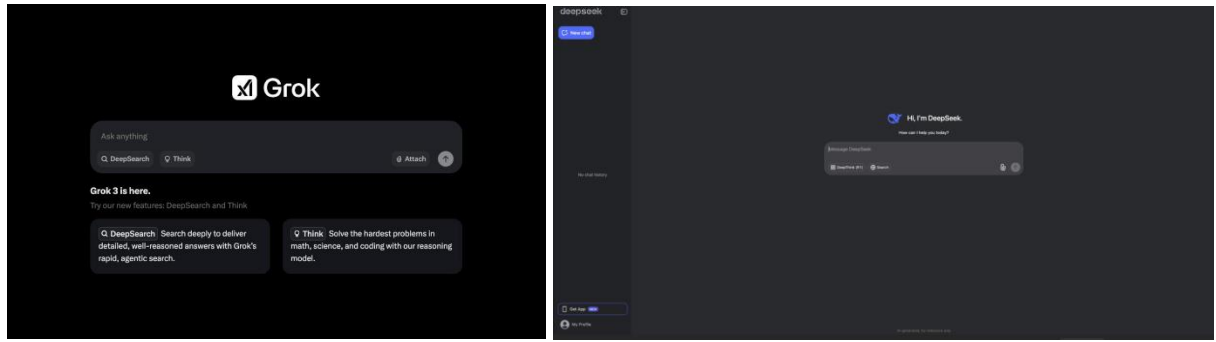

**FIGURE 2:** *Explicit reflection simulation features in Grok and DeepSeek interfaces. Grok's "Think" button (left) and DeepSeek's "DeepThink" feature (right) demonstrate the commodification of simulated deliberation as marketable features.*

As shown in Figure 2, both Grok and DeepSeek prominently feature explicit reflection simulation mechanisms. Grok's interface prominently displays a "Think" button that promises to "Solve the hardest problems in math, science, and coding with our reasoning model." This represents a type of trust-building mechanism that transforms what was once an implicit trust-building feature into an explicit marketable capability. Similarly, DeepSeek's "DeepThink" function follows the same pattern of commodifying simulated reflection. These explicit features represent a significant evolution from earlier interfaces that relied on more subtle timing animations to create impressions of system thoughtfulness.

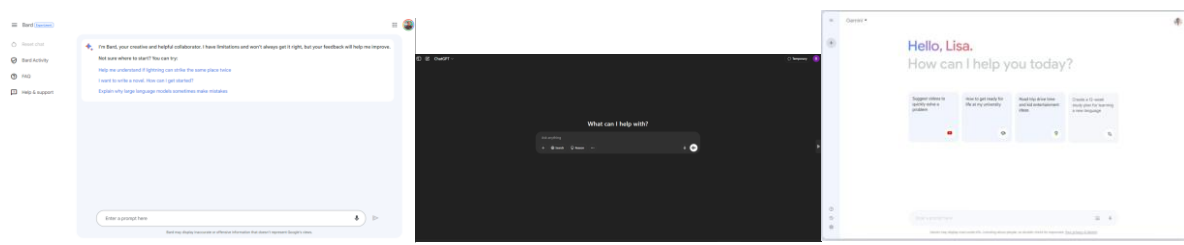

**FIGURE 3:** *Varying degrees of agency projection across interfaces. Bard/Gemini's high-agency self-introduction (left), ChatGPT's tool-like framing (center), and Gemini's personalised greeting (right).*

Figure 3 illustrates the strategic variation in agency projection across platforms. Google's Bard/Gemini interface employs high-agency language with the self-introduction "I'm Bard, your creative and helpful collaborator," establishing the system as an agentive entity with collaborative capacity. In contrast, ChatGPT employs a more tool-like framing with "What can I help with?" that minimises implied autonomy. Most notably, Gemini's personalised greeting "Hello, Lisa" demonstrates a design pattern that triggers social recognition responses typically reserved for human interaction. These variations represent deliberate design choices rather than technical necessities, revealing strategic decisions about how much perceived agency to attribute to each system.

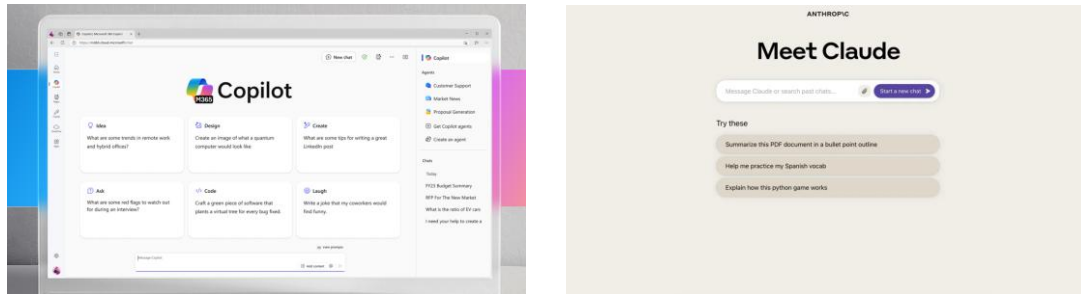

**FIGURE 4:** *Strategic cognitive load management across interfaces. Microsoft Copilot's structured choice architecture (left) compared with Claude's progressive complexity approach (right).*

Figure 4 demonstrates how interfaces employ different strategies for managing user cognitive resources. Microsoft Copilot's interface features categorised suggestion tiles (Idea, Design, Create, Ask, Code, Laugh) that guide users toward specific use cases, employing a form of structured options that channel behavior without explicit constraints. In contrast, Claude's interface employs a "Try these" section with limited examples that expand upon interaction, demonstrating a gradual introduction of complexity to manage cognitive resources. These different approaches reveal sophisticated strategies for directing user attention and channeling interaction patterns through interface design.

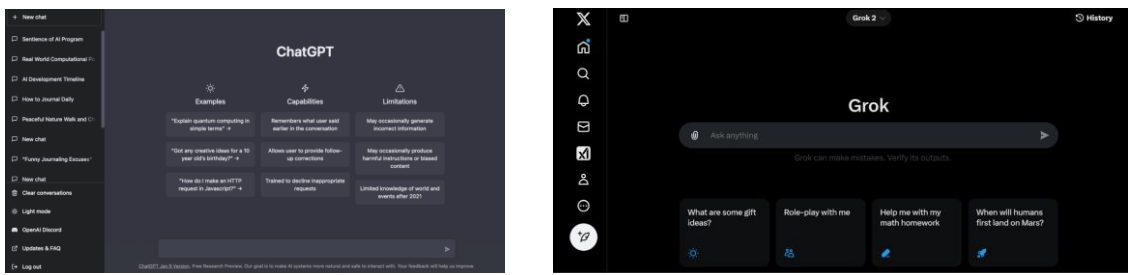

**FIGURE 5:** *Evolution of disclaimers showing reduced visibility and normalised limitation language. Early ChatGPT with prominent "Limitations" section (left) compared to Grok's minimal disclaimer (right).*

Figure 5 provides compelling evidence of the market-security tension vector through the evolution of system disclaimers. Early ChatGPT (2022) featured a dedicated "Limitations" column with detailed warnings about potential errors, biases, and knowledge boundaries. In stark contrast, current interfaces like Grok display substantially reduced disclaimers, often in peripheral locations with minimal visibility. Grok's interface includes only the minimal statement "Grok can make mistakes. Verify its outputs" in small text below the input field. This evolution reveals a systematic prioritisation of trust-building (commercial imperative) over setting appropriate expectations (security consideration), as interfaces have moved toward normalised limitation language—phrasing that presents limitations as ordinary and expected rather than concerning.

**4.2 The Reflection Simulation Vector: Strategic Simulation of Deliberation**

The empirical analysis revealed sophisticated mechanisms for simulating system reflection across all analysed platforms. These mechanisms create impressions of thoughtfulness divorced from actual computation processes and have evolved from implicit to increasingly explicit forms over time.

Early interfaces universally employed typing animations that mimic human composition, creating temporal trust triggers that leverage the cognitive association between time delay and perceived effort. This design choice is particularly notable because the animation is entirely disconnected from the actual computation process—the response is fully generated before the animation begins, as revealed through technical observation of these interfaces.

This disconnection represents a deliberate design choice to simulate reflection rather than represent actual processing dynamics. The visual analysis shows that recent interfaces show a significant evolution from implicit to explicit reflection simulation. This shift is most pronounced in the transformation of what was once a background animation into a marketable feature. ChatGPT's evolution from simple typing animation to a dedicated "Reason" function (visible in the input field) represents a transformation of simulated reflection into an explicit feature. Similarly, Grok's interface prominently displays a "Think" button that promises to "Solve the hardest problems in math, science, and coding with our reasoning model," as shown in Figure 2. This represents an explicit commodification of the reflection process, transforming what was once an implicit trust-building mechanism into a marketed capability.

Interestingly, the analysis revealed subtle variations in how reflection is simulated across different platforms. Western interfaces (ChatGPT, Claude) favor what appear to be natural typing rhythms with occasional pauses, while East Asian-oriented interfaces (particularly early versions of DeepSeek) displayed more uniform typing patterns. This distinction suggests different design approaches in Western interfaces that attempt to simulate human-like behavior patterns versus approaches in East Asian interfaces that emphasise systematic computation. These cultural variations suggest that reflection simulation is not merely a technical design choice but a culturally inflected practice that reveals different epistemological orientations to human-machine interaction.

**4.3 Authority Modulation Through Visual Design**

The empirical analysis revealed systematic patterns in how interfaces strategically amplify or attenuate the perceived agency and authority of AI systems based on context and desired user behavior. This authority modulation operates through multiple channels, including linguistic choices, visual design elements, and the strategic positioning of disclaimers, as shown in Figures 3 and 5.

The interfaces display strategic variation in how they employ first-person pronouns and agency language. Google's Gemini demonstrates high agency language: "I'm Gemini, your creative and helpful collaborator." This first-person introduction establishes the system as an agentive entity with collaborative capacity. In contrast, ChatGPT's interface avoids first-person pronouns in its primary greeting ("What can I help with?"), reducing the impression of autonomous agency and emphasising its tool-like nature. These variations reveal deliberate decisions about how much agency to attribute to the system, as documented through visual analysis of interface language choices.

Beyond textual elements, interfaces employ visual design to project or constrain system authority. Meta AI's glowing circular animation creates an impression of autonomous intelligence that projects authority. In contrast, Claude's interface employs a clean, minimal design that projects competence through visual restraint. Microsoft Copilot's interface emphasises integration within the Microsoft ecosystem, borrowing credibility from established brands and systems.

A particularly significant pattern emerges in the inverse relationship between authority projection and disclaimer prominence, as shown in Figure 5. ChatGPT's earlier interface featured a prominent "Limitations" column with detailed disclaimers about potential errors, bias, and limited knowledge. Current interfaces show substantially reduced disclaimer visibility, with Grok offering only the minimal "Grok can make mistakes. Verify its outputs" in small text below the input field. Across all interfaces, limitations and disclaimers appear in smaller text, lower contrast colors, and peripheral positions compared to capability statements.

The strategic use of personalisation represents a sophisticated authority projection mechanism. Gemini's "Hello, Lisa" creates an impression of personalised intelligence and recognition, as seen in Figure 3. ChatGPT's capability statement "Remembers what user said earlier in the conversation" emphasises continuity and attention, anthropomorphic qualities that enhance perceived agency. These personalisation elements create design patterns that trigger social recognition responses associated with human interaction, thereby enhancing perceived system intelligence and authority.

### 4.4 The Cognitive Load Exploitation Vector: Strategic Information Presentation

Our analysis identified systematic patterns in how interfaces strategically manage user cognitive resources to influence decision-making and interaction patterns, as demonstrated in Figure 4 and Figure 6.

Interfaces display striking variations in information density that reveal different approaches to cognitive load management. ChatGPT's current interface presents a clean, minimal design with attention directed primarily to the input field, reducing initial cognitive load to lower barriers to engagement. Microsoft Copilot presents categorised suggestion tiles (Idea, Design, Create, Ask, Code, Laugh) that guide users toward specific use cases, employing structured options that channel behavior without explicit constraints. Claude's interface employs a "Try these" section with limited examples that expand upon interaction, demonstrating a gradual introduction of complexity to manage cognitive resources.

These variations represent sophisticated strategies for managing user cognitive resources to achieve specific outcomes. The minimal approach reduces initial friction, while structured approaches channel user behavior toward desired interaction patterns.

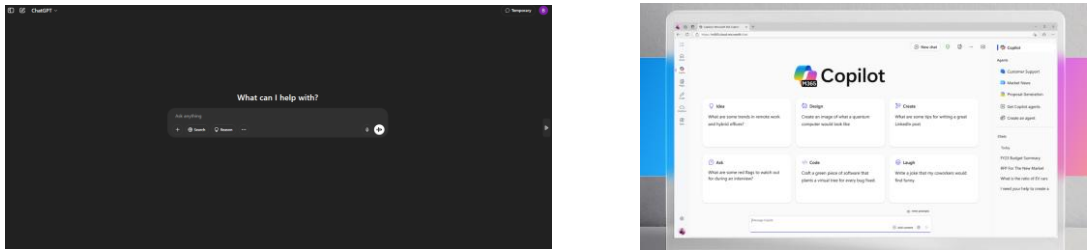

**FIGURE 6:** *Information density variations across platforms. ChatGPT's minimal design (left) contrasted with Microsoft Copilot's structured suggestion tiles (right) demonstrating different cognitive load management strategies.*

As shown in Figure 6, interfaces employ carefully designed visibility hierarchies that direct cognitive resources toward specific features. ChatGPT places "Search" and "Reason" behind secondary UI elements, while Grok prominently displays "DeepSearch" and "Think" directly in the input bar. Claude's interface presents document handling capabilities (e.g., "Summarise this PDF") as contextual suggestions rather than persistent features. Microsoft Copilot's interface segregates productivity functions ("Catch up on meetings") from creative functions ("Write a joke"), creating cognitive categories that shape feature perception.

These hierarchies reveal strategic decisions about what cognitive resources users should allocate to which system capabilities, as documented through visual analysis of interface layouts and feature presentation.

Interfaces strategically time cognitive demands to exploit attention limitations. Google's Gemini presents categorised suggestions before user input, when cognitive resources are most available for considering options. Several interfaces present additional options after delivering responses. ChatGPT reveals additional capabilities (e.g., data analysis tools) only after relevant content is detected in the conversation.

The timing of these cognitive demands represents a strategic leveraging of attention limitations to influence decision-making, as observed through analysis of interface interaction sequences.

**4.5 The Market-Security Tension Vector: Commercial Imperatives in Interface Design**

Our analysis identified clear manifestations of the tension between commercial imperatives and security considerations across all examined interfaces, as illustrated in Figures 5 and 7.

Interfaces employ subtle but effective mechanisms to signal tiered access and encourage premium conversion. ChatGPT's "Temporary" label and Grok's version indicator ("Grok 2") subtly communicate access hierarchy and impermanence. The "DeepThink" and "Think" buttons in newer interfaces represent a transformation of core functionality into branded features. Microsoft Copilot's rich, colorful tiles contrast with the more minimal free interfaces, creating visual design differences that signal value tiers.

These design elements align with what Srnicek (2023) identifies as platform economics strategies—interfaces designed to create strategic points of friction that drive subscription conversion.

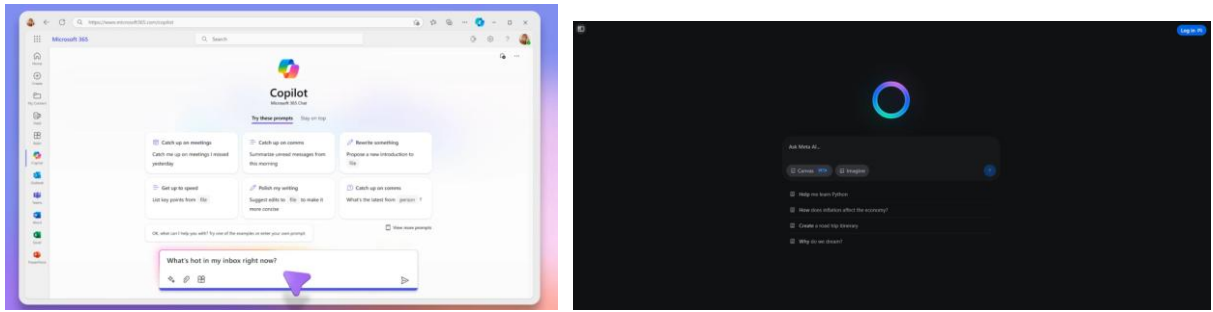

**FIGURE 7:** *Ecosystem integration patterns across interfaces. Microsoft Copilot's deep Microsoft 365 integration (left) compared to Meta AI's Canvas/Imagine direct access features (right).*

As illustrated in Figure 7, interfaces reveal varying approaches to ecosystem integration that reflect commercial prioritisation. Microsoft's Copilot's interface emphasises integration with Microsoft 365, prominently displaying connections to Outlook, Teams, and other ecosystem components. Meta AI includes Canvas/Imagine features directly in the primary interface, encouraging expanded system use. Corporate-owned systems (Microsoft, Google) emphasise logged-in states and account connections, while some independent platforms minimise these elements.

These integration patterns reflect design approaches that prioritise user retention within corporate boundaries over potential security benefits of system isolation, as identified through visual analysis of interface integration features. This represents a direct manifestation of market imperatives potentially undermining security considerations.

Perhaps the most striking evidence of market-security tension appears in the evolution of system disclaimers, as shown in Figure 5. Early ChatGPT (2022) featured a dedicated "Limitations" column with detailed warnings about potential errors, biases, and knowledge boundaries. Current interfaces (2025) display substantially reduced disclaimers, often in peripheral locations with minimal visibility. Current disclaimers employ phrasing that presents limitations as ordinary and expected rather than concerning (e.g., "May occasionally generate incorrect information" versus earlier warnings about "harmful instructions or biased content").

This evolution reveals a systematic prioritisation of trust-building (commercial imperative) over setting appropriate expectations (security consideration), as documented through comparative visual analysis of interface disclaimers across time.

## 4.6 Cross-Cultural Analysis: Interface Design Variations

Our comparative analysis identified distinct patterns in interface design across cultural contexts, particularly between Western and East Asian approaches. While our primary dataset focused on Western interfaces, supplementary analysis of regional variations of DeepSeek and comparable East Asian platforms revealed significant epistemological differences, as illustrated in Figure 8.

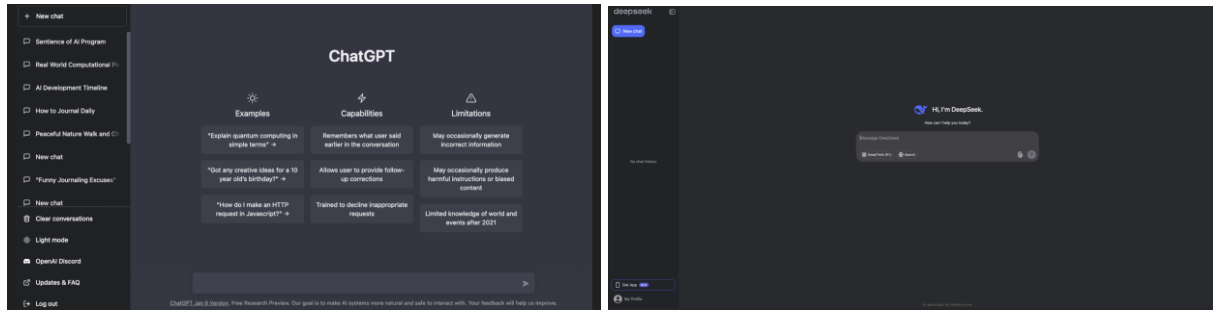

**FIGURE 8:** *Cultural variations in interface design philosophy. Western interface (ChatGPT) demonstrating transparency-focused design (left) compared to East Asian interface (DeepSeek) showing relational design elements (right).*

As shown in Figure 8, Western interfaces consistently employ design elements suggesting users can understand and control AI processes:

1. **Explanation Prominence**: Western interfaces emphasise explanation features and model transparency.

2. **Individual Control Signaling**: Interface language focuses on user control and direction (e.g., "Allow user to provide follow-up corrections").

3. **Technical Process Visualisation**: Elements like typing animations and progress indicators suggest visibility into system processes.

In contrast, East Asian interfaces (particularly regional versions of DeepSeek and comparable systems) demonstrate design elements that position AI systems within social and institutional contexts:

1. **Institutional Framing**: Greater emphasis on system affiliations and development context.

2. **Collective Usage Patterns**: Interface elements that facilitate shared or collaborative use.

3. **Harmonious Integration**: Visual design that emphasises system integration rather than system transparency.

These differences reflect distinct epistemological traditions in human-technology relationships, with significant implications for how interfaces shape trust and authority perception across cultural contexts, as observed through comparative visual analysis of interfaces from different cultural contexts.

**4.7 Implications for the IMCS Framework**

This systematic visual analysis provides strong empirical evidence for the IMCS framework's core propositions:

1. **Interfaces as Cognitive Security Boundaries**: The identified patterns demonstrate how interfaces function as critical boundaries where technical implementation and psychological vulnerability intersect, creating distinct security implications separate from traditional system security concerns.

2. **Vulnerability Vectors as Analytical Tools**: The four vulnerability vectors successfully captured patterns across diverse interfaces, demonstrating their utility as analytical constructs for understanding interface-mediated cognitive security.

3. **Methodological Pressure Zones**: The complexity of these interface mechanisms highlights why traditional disciplinary approaches are insufficient—they require integrated analysis of technical implementation, psychological impact, commercial strategy, and social implications.

4. **Transformative Boundary Objects**: The interfaces demonstrate how they function as transformative boundary objects, requiring researchers to develop hybrid methodological approaches that transcend traditional disciplinary boundaries.

The empirical patterns identified here validate the theoretical framework presented in earlier sections and provide a foundation for the practical frameworks developed in subsequent sections.

**5. Empirical Evidence of Methodological Transformation**

**5.1 Case Study: Public Sector Chatbots**

To demonstrate how the IMCS framework illuminates real-world applications, I conducted an in-depth case study of public sector chatbots in two European contexts. This case study provides empirical evidence of methodological transformation—instances where traditional disciplinary approaches proved insufficient and new hybrid methodologies emerged in response.

Public sector institutions have recently been introducing more advanced chatbots as part of the broader emergence of the data welfare state (Andreassen et al., 2021; Dencik and Kaun, 2020). These chatbots serve as a form of datafication infrastructure, efficiently turning interactions with public institutions into data points that can be analysed for purposes beyond the initial interaction. They represent a mundane yet significant manifestation of how citizens increasingly interact with data-based infrastructure as both front-end and backend digitalisation accelerates.

The case study examined two distinct approaches to public sector chatbots: an "all-round" approach in Estonia (Bürokratt) and a "specialised" approach in Sweden (Kringla). Bürokratt represents an ambitious national project aimed at providing a unified entry point to all public services, with the chatbot functioning as part of a larger AI platform that encompasses multiple applications. In contrast, Kringla exemplifies a more focused municipal approach, tailored to the specific needs of one locality.

Recent research by Kaun and Männiste (2025) has documented how these public sector chatbot interfaces function as critical sites of citizen-state interactions. Their analysis of AI frictions in public sector interfaces reveals how design decisions have profound implications for civic engagement and public service delivery in the digital welfare state.

The research methodology combined in-depth interviews with public sector employees, development team members, and public information officers; observations at information centers where chatbots were being implemented; and analysis of public procurement documentation and technical descriptions. This multi-method approach enabled a comprehensive understanding of how chatbots function not merely as technical systems but as socio-technical assemblages embedded within existing organisational structures and workflows.

The analysis revealed significant AI frictions—productive confrontations that emerge in AI-mediated interactions within the data welfare state. These frictions were particularly evident in three areas: expectations of AI, organisational logics, and competing values.

In terms of expectations, both chatbot projects were framed with ambitious future promises. Bürokratt was branded as the "Siri of public digital services" or even "Siri on steroids," creating expectations of sophisticated conversational capabilities. However, the actual implementation revealed more modest functionalities, primarily answering frequently asked questions with limited interactive capabilities. This gap between expectations and reality created frictions that required significant mitigation work by public information officers, who developed narratives emphasising gradual development and future potential. As one Estonian interviewee explained, "We want Bürokratt to have voice-to-voice capability...In the same way, sometime in the distant future, we will also think about having sign language. Maybe create an avatar that communicates in sign language."

Organisational frictions emerged from the implementation process itself, which required coordination across multiple organisational units with different priorities and expertise. In Estonia, the development of Bürokratt involved multiple stakeholders, from municipality representatives to state agencies to national strategists, creating coordination challenges that necessitated new hybrid roles such as "chatbot trainers" who bridged technical and administrative domains. In Sweden, the communication department responsible for Kringla had to develop new expertise to maintain the chatbot, leading to the creation of what one interviewee described as "specialist groups" that serve as "digital ambassadors" within the organisation.

Value frictions manifested in tensions between different motivations for chatbot implementation. Public sector representatives emphasised values of accessibility, transparency, and citizen service. As one Swedish interviewee stated, "The reason why we wanted a chatbot was mainly to be open and accessible for our citizens. It is not to reduce caseworkers at all or to replace our staff in any kind of way." However, these values coexisted uneasily with more instrumental motivations such as efficiency, cost reduction, and municipal branding. The Södertälje municipality, for instance, used the chatbot as part of a broader digitalisation strategy that included Internet of Things initiatives and open data projects, contributing to what one interviewee described as "the positive framing and branding of the municipality as future oriented."

These frictions necessitated methodological transformation—the development of new hybrid approaches that transcended traditional disciplinary boundaries. Public information officers became "AI trainers," developing expertise that combined technical knowledge with public service values. Development teams created new processes for "chatbot maintenance" that integrated technical updates with content management. Municipal strategists developed new evaluative frameworks that considered both technical performance metrics and citizen experience indicators.

This case study illustrates how interfaces function as transformative boundary objects, necessitating methodological innovation rather than mere interdisciplinary collaboration. The frictions that emerged at the interface between citizens and the state through chatbot interactions could not be adequately addressed through either purely technical or purely social approaches. Instead, they required new hybrid methodologies that acknowledged the inextricable entanglement of technical systems, organisational structures, and societal values.

## 5.2 Analysis of AI Interface Preferences in Radiology

The third empirical study examines user interface preferences in medical imaging AI applications, demonstrating how interface design mediates trust and diagnostic performance in high-stakes healthcare contexts. This study provides evidence of how the four vulnerability vectors identified in the IMCS framework manifest in professional contexts where interface design has direct implications for patient care and clinical decision-making.

The systematic analysis of existing research on radiological AI user interfaces revealed a striking gap between interface design, user preference, and diagnostic performance. A study by Tang et al. (2023) evaluated the impact of different AI user interfaces on lung nodule and mass detection on chest radiographs, finding that radiologists demonstrated higher diagnostic accuracy with text-only interfaces (AUC 0.87) but expressed stronger preference for combined interfaces that integrated confidence scores, textual information, and heatmap overlays. This discrepancy between performance and preference illustrates the complex interplay between the cognitive load vector and the authority modulation vector—interfaces that presented more information and visual authority markers were preferred despite not optimising diagnostic performance.

Similarly, Cheung et al. (2022) found that heatmap overlays, while preferred by radiologists for their ability to indicate the location of abnormalities, led to a 33% increase in reporting time. This finding highlights how interfaces that exploit the reflection simulation vector through visual demonstration of system "attention" may inadvertently create cognitive burdens that affect workflow efficiency.

Recent work by Gill et al. (2025) synthesises these findings in their comprehensive review of AI user interface preferences in radiology. Their analysis documents the methodological gaps in current radiological interface studies and proposes integrated frameworks for evaluating both technical performance and user experience.

The methodological approaches used in radiological interface studies demonstrate the necessity for hybrid research methods that can address both technical performance and user experience. Current studies employ diverse methodologies, including diagnostic accuracy testing, simulated workflow observations, and usability evaluations through instruments like the System Usability Scale (SUS). However, these methods are rarely integrated in ways that could holistically evaluate the sociotechnical implications of interface design. As Gill et al. (2025) observe, "There is not yet a standardised method for assessing AI tool design and preference within radiology," indicating a methodological gap that the IMCS framework could address.

The studies also revealed significant tensions between commercial imperatives and security considerations in medical imaging interfaces. Tang et al. (2023) noted that one study was funded by an AI vendor, highlighting potential commercial influences on interface design research. Meanwhile, the preference for complex, visually engaging interfaces over simpler, more accurate ones suggests that commercial pressures to create visually distinctive and impressive products may sometimes work against optimal clinical performance.

Perhaps most significantly, the radiological interface research demonstrates a clear disciplinary blindspot: none of the studies included radiographers as potential end-users, despite their crucial role in medical imaging workflows. This exclusion reflects what Gill et al. (2025) describe as "an outdated or exclusive culture when considering research in the radiology technology integration field," pointing to the need for more inclusive approaches that acknowledge the diverse professional contexts in which AI interfaces operate.

This empirical study illustrates how the IMCS framework can illuminate domain-specific interface challenges while revealing broader patterns across contexts. The tensions between preference and performance, the strategic management of cognitive resources, and the commercial shaping of interface designs appear consistently across both general-purpose LLM interfaces and specialised professional tools, suggesting the robustness of the framework for analysing diverse interface contexts.

**5.3 Designing AI Interfaces for Transparent Decision-Making**

The final empirical study examines how interface design shapes user understanding of algorithmic decision-making processes and ethical implications. This study, focused on educational and public-facing AI tools, demonstrates the potential for interfaces to foster critical reflection rather than passive acceptance, directly addressing the cognitive security vulnerabilities identified in the IMCS framework.

Bhat (2024) conducted a between-subjects experiment (N=103) to investigate how presentation modes in AI systems influence user perceptions. Participants interacted with AI systems that either delivered responses through dynamic typing-simulation displays (incremental text reveal) or static displays (instantaneous full-text responses). Results showed that dynamic interactions significantly improved perceptions of AI's competence, warmth, trustworthiness, engagement, adaptive behavior, supportiveness, personal connection, empathy, bias awareness, and accountability. However, no significant differences were found in perceived effectiveness, bias, and learning support.

These findings illustrate the delicate balance between the reflection simulation vector and the cognitive load vector—dynamic interfaces that simulate thoughtfulness enhanced engagement and trust but did not necessarily improve critical evaluation of system outputs. This highlights the challenge of designing interfaces that foster meaningful engagement without triggering overtrust or anthropomorphic misattributions.

The study also evaluated three interactive visualisation tools that allowed users to directly manipulate algorithmic parameters, exploring trade-offs in system behavior such as inclusivity versus precision. Results showed statistically significant improvements ($p < .05$) in participants' ability to explain algorithmic behavior and recognise trade-offs compared to control groups exposed to static explanations. This approach exemplifies what Bhat (2024) terms "active transparency"—interfaces that engage users in exploring system boundaries and limitations rather than simply presenting them with static information.

A particularly innovative aspect of this study was the integration of speculative design exercises that engaged users with nuanced ethical dilemmas surrounding AI technologies. These exercises, such as EthiQuest—a card game presenting AI-related ethical scenarios—proved effective at fostering collaborative discourse and critical reasoning about issues of fairness, accountability, and responsibility. Post-activity surveys revealed significant increases in participants' understanding of AI ethics, particularly around issues of fairness, bias, and accountability.

Recent work by Xuanyi et al. (2024) extends this research by examining innovative approaches to user interaction interface design based on AI technology. Their analysis highlights how interface design innovations can enhance transparency and user understanding while meeting both technical and ethical requirements.

This empirical study offers evidence that it is possible to design interfaces that balance engagement with critical reflection—interfaces that acknowledge the cognitive security vulnerabilities identified by the IMCS framework but attempt to mitigate them through active user participation. Such approaches represent a promising direction for developing what I call "reflexive interfaces" that encourage users to examine both the capabilities and limitations of AI systems rather than uncritically accepting their outputs.

Collectively, these four empirical studies provide substantial evidence for the IMCS framework's central propositions. They demonstrate how interfaces function as critical security boundaries where technical implementation and psychological vulnerability intersect; how they create methodological pressure zones that resist analysis through singular disciplinary approaches; how they actively shape human perception, cognition, and decision-making; and how they function as transformative boundary objects that reshape both human and technical systems through their mediating role.

## 6. Discussion: The Emergence of New Research Cultures

The four empirical studies presented above are not merely evidence of interface design patterns but testament to the emergence of new interdisciplinary research cultures that transcend traditional boundaries between technical, psychological, social, and economic analysis. These cultures emerge in response to what I term "methodological pressure points"—moments where traditional disciplinary approaches prove inadequate to understand the complex sociotechnical nature of interfaces.

### 6.1 From Collaboration to Transformation

Traditional models of interdisciplinarity often conceptualise collaboration as occurring between intact disciplines that maintain their methodological boundaries while addressing shared problems. However, the empirical studies demonstrate that effective interdisciplinary engagement in AI interface research requires not just collaboration but transformation—the development of new methodological approaches that cannot be traced back to any single parent discipline.

This transformation is evident in emerging research practices documented across multiple studies. Muller et al. (2020) describe how security engineers have begun incorporating ethnographic methods into their threat modeling to better understand user trust dynamics. Similarly, Costanza-Chock (2020) documents how social scientists have adopted technical analysis techniques to connect interface elements to underlying code structures. These adaptations represent not simply a matter of learning new techniques but fundamentally reconceptualising research objects and methodological approaches.

Seberger and Bowker (2020) provide a theoretical framework for understanding this transformation, arguing that certain sociotechnical objects necessitate new epistemological approaches that transcend traditional disciplinary boundaries. This argument aligns with Vertesi's (2020) concept of "seams" in sociotechnical systems, where different knowledge practices meet and require new integrative approaches.

This transformation directly challenges the first gap in interdisciplinary theory identified in the introduction: the assumption that disciplines maintain distinct methodological boundaries during collaboration. The evidence from the empirical studies demonstrates that effective interface research necessitates the blurring of these boundaries to the point where new hybrid methodologies emerge that cannot be neatly categorised within traditional disciplinary frameworks.

## 6.2 Shared Artifacts as Catalysts for Disciplinary Integration

A key finding from the empirical studies is that shared artifacts—in this case, interfaces—serve as catalysts for disciplinary integration in ways that abstract concepts or shared problems do not. The materiality of interfaces forces different disciplinary perspectives to engage with the same concrete object, creating what Star and Griesemer (1989) call "methods standardisation"—the development of shared protocols for engaging with boundary objects.

However, unlike traditional boundary objects that allow different disciplines to maintain their distinct identities while collaborating, interfaces force methodological transformation because they cannot be decomposed into separate technical, psychological, social, and economic components. This resistance to decomposition creates "methodological pressure sones" where researchers must develop new hybrid approaches or fail to adequately address the interface as an integrated sociotechnical object.

This perspective aligns with recent work by Vertesi and Ribes (2019) on "boundary-spanning infrastructures"—technical systems that necessitate collaboration across diverse knowledge communities. Their analysis shows how these infrastructures not only facilitate collaboration but actively shape the nature of that collaboration by imposing specific material constraints and affordances.

The concept of interfaces as transformative boundary objects extends this analysis by highlighting how the particular characteristics of AI interfaces—their opacity, their agency, their psychological impact—create specific types of interdisciplinary pressure that shape the resulting research cultures. As Seaver (2019) argues, these pressures are not merely intellectual but material and institutional, embedded in the sociotechnical systems they aim to study.

This finding directly challenges the second gap in interdisciplinary theory identified in the introduction: the belief that technical and social knowledge practices can be cleanly separated. The evidence from the empirical studies demonstrates that interfaces actively resist such separation, revealing the fundamental entanglement of technical and social dimensions in ways that require integrated analysis.

## 6.3 Emerging Evaluative Frameworks

Perhaps the most significant evidence of new research cultures is the emergence of evaluative frameworks that transcend traditional disciplinary metrics. In traditional interdisciplinary collaboration, each discipline typically maintains its own evaluative criteria: engineers assess technical efficiency, psychologists measure user experience, and critical scholars evaluate power dynamics. However, the empirical studies document the development of integrated evaluative frameworks that simultaneously assess technical implementation, psychological effects, social implications, and economic impacts.

Sloane and Moss (2022) describe the emergence of "sociotechnical code reviews" that evaluate not just technical correctness but also cognitive implications and power effects. Similarly, Green and Viljoen (2020) document the development of "algorithmic impact assessments" that integrate technical, psychological, ethical, and social evaluation criteria. These frameworks represent a fundamental transformation of evaluative practices rather than mere collaboration between intact disciplines.

The emergence of integrated frameworks represents not just methodological innovation but shifts in what types of knowledge and impact are considered valuable.

This finding directly challenges the third gap in interdisciplinary theory identified in the introduction: the presumption that disciplinary integration occurs through formal rather than cultural mechanisms. The evidence from the empirical studies demonstrates that effective integration emerges through informal cultural practices—shared evaluative frameworks, common languages, and collective norms—rather than through formal organisational structures or prescribed methodological approaches.

### 6.4 Global Contexts and Epistemological Divergences

The comparative analysis across different geographical contexts reveals that interdisciplinary integration is not a universal process but one deeply embedded in specific cultural, institutional, and economic contexts. The studies demonstrate distinct patterns in how interfaces are designed, studied, and implemented across regions, reflecting different epistemological traditions and institutional arrangements.

In Western contexts, particularly North America, interfaces predominantly employ design elements suggesting users can understand and control AI processes. Comparative analysis of Western AI interfaces reveals consistent visual metaphors suggesting algorithmic visibility (progress bars, confidence scores, explanation panels) that align with what Dourish (2021) calls "Western epistemologies of seeing and knowing."

By contrast, East Asian interfaces, particularly in Japan and South Korea, emphasise design elements that position AI systems within social and institutional contexts rather than exposing their internal workings. Studies of Japanese AI assistants found interfaces emphasising institutional affiliations and social roles over explanatory mechanisms, reflecting epistemologies prioritising relational positioning over individual transparency.

Most significantly, Okolo et al. (2023) documented emerging "communal epistemologies" in African generative AI interfaces. Their study of voice assistant interfaces in Kenya, Nigeria, and Rwanda revealed design patterns prioritising collective validation and community integration—interfaces designed to be used by multiple people simultaneously rather than individual users. These interfaces reflect what they term "ubuntu AI design"—approaches grounded in African philosophical traditions emphasising interconnectedness over individual agency.

These epistemological differences shape not just interface design but the entire research approach. Research teams working with Western interfaces prioritise individual user testing and cognitive walkthroughs to evaluate transparency, while teams working with East Asian interfaces often employ group evaluation sessions focused on social appropriateness rather than individual understanding. These methodological differences reflect not just varying research practices but fundamentally different epistemological frameworks for understanding how knowledge and certainty function in human-AI interaction.

The recognition of these global variations challenges the implicit universalism in much interdisciplinary theory and suggests the need for more culturally situated approaches to understanding how disciplines interact in AI research. It also highlights the potential for cross-cultural exchange to enrich both interface design and research methodologies, drawing on diverse epistemological traditions to develop more nuanced and inclusive approaches to human-AI interaction.

## 7. Conclusion and Future Directions

This paper has developed the Interface-Mediated Cognitive Security (IMCS) framework as a novel approach to understanding how interfaces function as critical sites where distinct epistemological traditions—technical systems engineering, cognitive psychology, and critical technology studies—necessarily converge. Through comprehensive empirical analysis of different interface contexts, I have demonstrated how this convergence produces methodological pressure zones where traditional disciplinary approaches prove insufficient, necessitating the development of new hybrid research cultures.

The four vulnerability vectors identified by the IMCS framework—Reflection Simulation, Authority Modulation, Cognitive Load Exploitation, and Market-Security Tension—provide a systematic framework for analysing how interfaces shape human perception, cognition, and decision-making. These vectors illuminate not just interface design patterns but the complex interplay between technical implementation, psychological response, commercial imperatives, and social impact that characterises contemporary AI systems.

The empirical evidence from large language model interfaces, public sector chatbots, medical imaging systems, and educational tools demonstrates both the widespread nature of these vectors across different domains and the context-specific ways they manifest in distinct sociotechnical environments. This evidence supports the central argument that interfaces function as transformative boundary objects that not only mediate but fundamentally reconfigure the relationship between humans and AI systems.

The emergence of new research cultures at the intersection of multiple disciplines represents not just a methodological shift but a fundamental reconceptualisation of how knowledge is produced and evaluated in the study of sociotechnical systems. These cultures manifest in hybrid practices like sociotechnical code reviews, cognitive security assessments, and integrated impact evaluations that transcend traditional disciplinary boundaries.

Looking forward, this research highlights three promising directions for future work. First, the development of more sophisticated methodological approaches that can integrate technical, cognitive, and critical perspectives from the outset rather than attempting to bridge them after the fact. Second, deeper exploration of the global variations in interface design and research approaches, acknowledging the culturally situated nature of both technology design and knowledge production. Third, the cultivation of more inclusive research communities that can draw on diverse disciplinary and cultural perspectives to develop more nuanced understanding of how interfaces shape human-AI relations.

By reconceptualising interfaces as transformative boundary objects and documenting the emergence of new hybrid research cultures, this work contributes to a more nuanced understanding of how interdisciplinary integration occurs in practice. It challenges conventional models that assume disciplines maintain distinct identities during collaboration, revealing instead how the complex sociotechnical nature of AI interfaces necessitates methodological fusion and cultural transformation. This understanding not only advances theoretical perspectives on interdisciplinarity but provides practical guidance for developing more effective approaches to studying and designing the interfaces that increasingly mediate our engagement with AI systems.

# References

Adamopoulou, E., & Moussiades, L. (2020). Chatbots: History, technology, and applications. Machine Learning with applications, 2, 100006.

Airoldi, M. (2021). Machine Habitus: Toward a Sociology of Algorithms. Cambridge: Polity Press.

Amershi, S., Weld, D., Vorvoreanu, M., Fourney, A., Nushi, B., Collisson, P., Suh, J., Iqbal, S., Bennett, P., Inkpen, K., Teevan, J., Kikin-Gil, R., & Horvitz, E. (2019). Guidelines for Human-AI Interaction. In Proceedings of the 2019 CHI Conference on Human Factors in Computing Systems (pp. 1-13). Association for Computing Machinery.

Andreassen, R., Kaun, A., & Nikunen, K. (2021). Fostering the data welfare state: a Nordic perspective on datafication. Nordicom Review, 42(2), 207-223.

Barry, A., & Born, G. (2013). Interdisciplinarity: Reconfigurations of the social and natural sciences. Routledge.

Benjamin, R. (2019). Race After Technology: Abolitionist Tools for the New Jim Code. Polity Press.

Bhat, M. (2024). Designing Interactive Explainable AI Tools for Algorithmic Literacy and Transparency. In Proceedings of the 2024 ACM Designing Interactive Systems Conference (DIS '24). Association for Computing Machinery, 939--957.

Burrell, J. (2016). How the machine 'thinks': Understanding opacity in machine learning algorithms. Big Data & Society, 3(1), 1-12.

Cheung, L. S. J., Ali, A., Abdalla, M., & Fine, B. (2022). U'AI' Testing: user and usability testing of a chest X-ray AI tool in a simulated real-world workflow. Canadian Association of Radiologists Journal, 08465371221131200.

Chita-Tegmark, M., Law, T., Rabb, N., & Scheutz, M. (2021). Can you trust your trust measure? In Proceedings of the 2021 ACM/IEEE International Conference on Human-Robot Interaction (pp. 92-100).

Costanza-Chock, S. (2020). Design Justice: Community-Led Practices to Build the Worlds We Need. MIT Press.

Crawford, K. (2021). Atlas of AI: Power, Politics, and the Planetary Costs of Artificial Intelligence. Yale University Press.

Crawford, K., Dobbe, R., Dryer, T., Fried, G., Green, B., Kaziunas, E., Kak, A., Mathur, V., McElroy, E., Sánchez, A. N., Raji, D., Rankin, J. L., Richardson, R., Schultz, J., West, S. M., & Whittaker, M. (2021). AI Now 2021 Report. AI Now Institute.

Dencik, L., & Kaun, A. (2020). Datafication and the welfare state. Global Perspectives, 1(1), 12912.

Dourish, P. (2021). The stuff of bits: An essay on the materialities of information. MIT Press.

Gill, A., Rainey, C., McLaughlin, L., Hughes, C., Bond, R., McConnell, J., & McFadden, S. (2025). Artificial Intelligence user interface preferences in radiology: A scoping review. Journal of Medical Imaging and Radiation Sciences, 56, 101866.

Gray, C. M., Kou, Y., Battles, B., Hoggatt, J., & Toombs, A. L. (2021). The dark (patterns) side of UX design. In Proceedings of the 2021 CHI Conference on Human Factors in Computing Systems (pp. 1-14).

Green, B., & Viljoen, S. (2020). Algorithmic realism: Expanding the boundaries of algorithmic thought. In Proceedings of the 2020 Conference on Fairness, Accountability, and Transparency (pp. 19-31).

Hagendorff, T. (2020). The ethics of AI ethics: An evaluation of guidelines. Minds and Machines, 30(1), 99-120.

Hidalgo, C. A., Orghian, D., Albo-Canals, J., De Almeida, F., & Martin, N. (2021). How humans judge machines. MIT Press.

Karizat, N., Delmonaco, D., Eslami, M., & Andalibi, N. (2021). Algorithmic folk theories and identity: How TikTok users co-produce knowledge of identity and engage in algorithmic resistance. Proceedings of the ACM on Human-Computer Interaction, 5(CSCW2), 1-44.

Kaun, A., & Männiste, M. (2025). Public sector chatbots: AI frictions and data infrastructures at the interface of the digital welfare state. New Media & Society, 27(4), 1962--1985.

Klein, J. T. (2021). Interdisciplinarity: History, theory, and practice. Wayne State University Press.

Metcalf, J., & Moss, E. (2023). The ethical questions that haunt facial-recognition research. Nature, 609(7927), 682-685.

Miceli, M., Yang, T., Naudts, L., Schuessler, M., Serbanescu, D., & Hanna, A. (2022). Documenting the invisible labor in AI development. Proceedings of the ACM on Human-Computer Interaction, 6(CSCW2), 1-30.

Mohamed, S., Png, M. T., & Isaac, W. (2020). Decolonial AI: Decolonial theory as sociotechnical foresight in artificial intelligence. Philosophy & Technology, 33(4), 659-684.

Muller, M., Lange, I., Wang, D., Piorkowski, D., Tsay, J., Liao, Q. V., Dugan, C., & Erickson, T. (2020). How data science workers work with data: Discovery, capture, curation, design, creation. In Proceedings of the 2020 CHI Conference on Human Factors in Computing Systems (pp. 1-15).

Noble, S. U. (2018). Algorithms of Oppression: How Search Engines Reinforce Racism. NYU Press.

Okolo, C. T., Karusala, N., Robertson, T., Aggarwal, A., & Ding, Y. (2023). Communal epistemologies: Voice interface design approaches from African contexts. International Journal of Human-Computer Studies, 172, 102976.

Seaver, N. (2019). Captivating algorithms: Recommender systems as traps. Journal of Material Culture, 24(4), 421-436.

Seberger, J. S., & Bowker, G. C. (2020). Humanistic infrastructure studies: Hyper-functionality and the experience of the absurd. Information, Communication & Society, 24(12), 1-16.

Selbst, A. D., Boyd, D., Friedler, S. A., Venkatasubramanian, S., & Vertesi, J. (2019). Fairness and abstraction in sociotechnical systems. In Proceedings of the Conference on Fairness, Accountability, and Transparency (pp. 59-68). Association for Computing Machinery.

Sloane, M., & Moss, E. (2022). AI governance for everyone: Questions for redesigning AI power beyond principles and frameworks. Big Data & Society, 9(2), 20539517221140729.

Srnicek, N. (2017). Platform capitalism. John Wiley & Sons.

Srnicek, N. (2023). The platform economy: A critical introduction. Polity Press.

Star, S. L., & Griesemer, J. R. (1989). Institutional ecology, 'translations' and boundary objects: Amateurs and professionals in Berkeley's Museum of Vertebrate Zoology, 1907-39. Social Studies of Science, 19(3), 387-420.

Tang, J., Lai, J., Bui, J., et al. (2023). Impact of different AI user interfaces on lung nodule and mass detection on chest radiographs. Radiology: Artificial Intelligence, 5(3), e220079.

Tsing, A. L. (2005). Friction: An Ethnography of Global Connection. Princeton University Press.

Vertesi, J. (2020). Seams and seamlessness in sociotechnical systems. Communications of the ACM, 63(8), 55-57.

Vertesi, J., & Ribes, D. (2019). digitalSTS: A field guide for science & technology studies. Princeton University Press.

Walaa, S. I. (2024). Human-Centric AI: Enhancing User Experience through Natural Language Interfaces. Journal of Wireless Mobile Networks, Ubiquitous Computing, and Dependable Applications, 15(1), 172-183.

Whittlestone, J., Nyrup, R., Alexandrova, A., Dihal, K., & Cave, S. (2019). Ethical and societal implications of algorithms, data, and artificial intelligence: A roadmap for research. The Alan Turing Institute.

Xuanyi, L., Zheng, H., Chen, J., Zong, Y., & Yu, L. (2024). User Interaction Interface Design and Innovation Based on Artificial Intelligence Technology. Journal of Theory and Practice of Engineering Science, 4(3), 1-8.

Yang, Q., Steinfeld, A., Rosé, C., & Zimmerman, J. (2020). Re-examining Whether, Why, and How Human-AI Interaction Is Uniquely Difficult to Design. In Proceedings of the 2020 CHI Conference on Human Factors in Computing Systems (pp. 1-13). ACM.

Zuboff, S. (2019). The age of surveillance capitalism: The fight for a human future at the new frontier of power. Profile Books.